\documentclass[10pt]{amsart}
     \usepackage[german,english]{babel}
     \usepackage{enumerate}
     \usepackage{exscale}
     \usepackage{amsmath,amsfonts,amssymb}
     \newcommand{\be}{\begin{equation}}
     \newcommand{\ee}{\end{equation}}
     \newcommand{\bea}{\begin{eqnarray*}}
     \newcommand{\eea}{\end{eqnarray*}}
     \newcommand{\beq}{\begin{eqnarray}}
     \newcommand{\eeq}{\end{eqnarray}}
     \newcommand{\RR}{\mathbb{R}}
     \newcommand{\NN}{\mathbb{N}}
     \newcommand{\ZZ}{\mathbb{Z}}
     \newcommand{\EE}{\mathbb{E}}
     \newcommand{\PP}{\mathbb{P}}
     \newcommand{\cA}{\mathcal{A}}
     \newcommand{\cB}{\mathcal{B}}
     \newcommand{\cF}{\mathcal{F}}
     
     \newcommand{\cH}{\mathcal{H}}

     \newcommand{\tR}{\tilde R}
     \renewcommand{\r}{\right}
     \renewcommand{\l}{\left}
     \newcommand{\la}{\langle}
     \newcommand{\ra}{\rangle}
     
     \newcommand{\nr}[1]{\vert #1\vert}

     \newcommand{\Tr}{\mathop{\mathrm{Tr}}}
     \newcommand{\dist}{\mathrm{dist}}
     \newcommand{\proj}{\mathrm{proj}}

     \newtheorem{thm}{Theorem}[section]
     
     \newtheorem{prp}[thm]{Proposition}
     
     \theoremstyle{definition}

     \theoremstyle{remark}

\begin{document}

\title{Quantum site percolation on amenable graphs}

\author{Ivan Veseli\'c}
\address{Fakult\"at f\"ur Mathematik, D-09107 TU Chemnitz }
\email{ivan.veselic@mathematik.tu-chemnitz.de}
\urladdr{www.tu-chemnitz.de/mathematik/schroedinger/}

\keywords{integrated density of states, random Schr\"odinger operators,
random graphs, site percolation}
\subjclass[2000]{35J10,81Q10,82B43}

\begin{abstract}
We consider the quantum site percolation model on graphs with an amenable group
action. It consists of a
random family of Hamiltonians.
Basic spectral properties of these operators are derived: non-randomness
of the spectrum and its components, existence
of an self-averaging integrated density of states and an associated
trace-formula.
\end{abstract}

\thanks{\today}
\maketitle

\section{Introduction: The Quantum percolation model}

The quantum percolation model (QPM) consist of two building blocks which are
both well studied in physics of disordered media.
\smallskip

Let us first introduce the \emph{classical site percolation model}.
It is used to model the flow of liquid through porous
media, the spreading of forest fires or of
diseases etc.
Consider the graph $\ZZ^d$ where two vertices are connected by an
edge if their Euclidean distance equals one. Equip each vertex $v
\in \ZZ^d$ with a random variable $q(v)$ taking the value $0$ with
probability $p$ and "$\infty$" with probability $1-p$ and being
independent of the random variables at all other vertices. For
each configuration of the randomness $\omega \in \Omega$ let
$V(\omega):= \{v \in \ZZ^d \mid q(v)=0\}$.
The percolation problem
consists in studying the properties of connected components ---
called \emph{clusters} --- of $V(\omega)$. Typical questions are: With
what probability do infinite clusters exist? What is the average
vertex number or diameter of a cluster? What is the probability
that $0,v \in \ZZ^d$ are in the same cluster, etc.?
One of the
central results of percolation theory is the existence of a
critical probability  $p_c$, such that for $p> p_c$ (respectively
for $p< p_c$) an infinite cluster exists (respectively does not
exist) almost surely. See e.g.~\cite{Kesten-82,Grimmett-99} and
the literature cited there.
\smallskip

\emph{Random lattice Hamiltonians} are used to describe the motion of waves in disordered media.
Each of them is a family of operators on $l^2(\ZZ^d)$ indexed by elements of a probability space.
The family obeys an equivariance relation with respect to the action of a group. More precisely, 
the group acts  in a consistent way on $l^2(\ZZ^d)$ by translations and on the probability space by ergodic, measure preserving transformations.

The spectral features of these random operators allow one to draw conclusions
about the transport and wave-spreading properties of the modelled medium.
Monograph expositions of this topic can be found in
\cite{CarmonaL-1990,PasturF-92,Stollmann-01}.
\smallskip

Let us define the simplest QPM: Let
\be
\label{e-QPM}
(A_\omega f)(v)
= \hspace{-.3em} \sum_{\dist(v,w)=1} \hspace{-.3em}  f(w) \quad
\text{ for all $v,w\in V(\omega)$ and all $f\in l^2(V(\omega))$}
\ee
be the adjacency operator of $V(\omega)$ introduced above. More precisely, $A_\omega$ is the adjacency operator of the induced sub-graph of $\ZZ^d$ with vertex set $V(\omega)$.
Here "$\dist$" denotes the distance function on this graph.

At this point let us explain why we chose $\infty$ as one of the values the random variable $q(v)$ takes.
The adjacency operator on $\ZZ^d$ corresponds (up to an energy shift) to the kinetic energy part of a quantum Hamiltonian on the lattice.
In this picture $q$ corresponds to the potential energy. In the quantum percolation model it vanishes on some sites, on others it is infinitely high, i.e.~forms an impenetrable barrier for the quantum wave function.

The interesting feature of the QPM is that it defines a Laplacian on
random geometry. More precisely, its domain of definition $l^2(V(\omega))$ varies with
$\omega$. This is the main difference to the random lattice operators considered in
\cite{CarmonaL-1990,PasturF-92}.
After an extension of the notion of random lattice Hamiltonians the QPM fits in this framework.
In our approach we rely on methods from \cite{PeyerimhoffV-2002,LenzPV-2002,LenzPV-2002?},
developed there to study operators on manifolds.

The QPM was first studied in \cite{deGennesLM-59a,deGennesLM-59b}.
There it was considered as a quantum mechanical model
for electron-propagation in binary alloys
where only one of the two types of atoms participates in the spreading of
the wavepacket.
The model attracted special attention because of
the existence of \emph{molecular states}, i.e.~eigenvectors supported on
finite regions of the infinite cluster, see
\cite{KirkpatrickE-72,ChayesCFJS-86}. The last cited reference is the
motivation of the present paper and our results can be seen as a
mathematically rigorous version of some arguments in \cite{ChayesCFJS-86}
and their extension to more general graphs.
\smallskip

The integrated density of states (IDS) of a Hamiltonian is the number of
eigenvalues per unit volume below a certain energy value. Thanks to the
stationarity and ergodicity assumptions it is well defined for random
Hamiltonians.
The IDS of a random Hamiltonian captures its global spectral features and
its understanding is the prerequisite of the study of finer spectral
properties. In the present work we analyze this quantity and provide
therewith a basis for a further study of the QPM, cf.~Section
\ref{s-disc} and \cite{Veselic-QP2}.

The next section states the results of this note, Section \ref{s-thm} is
devoted to their proofs and the last section concludes with a discussion of
further research topics.

\section{Results: Spectral properties of finite range hopping operators}
\label{s-thm}
To describe the geometric setting we are working in precisely, let us
recall basic notions from graph theory and  fix the notation along the
way. A \emph{graph} $G=(V,E)$ is given by a set of \emph{vertices}
$V=V(G)$ and a set of \emph{edges} $E=E(G) \subset \big (V\times V\setminus
\{(v,v)\mid  v \in V\}\big )/\sim$. Here $\sim$ denotes the relation
$(v,w)\sim(w,v)$. If $e= (v,w)\in E$, we call $v,w\in V$ \emph{nearest neighbours}
and \emph{endpoints} of the edge $e$. By our definition a
graph is \emph{simple}: it contains neither multiple edges nor self-loops,
i.e.~edges joining a vertex to itself.

A \emph{path} (of length $n$) in $G$ is an alternating sequence of vertices and edges
$\{v_0, e_1, v_1, \dots e_n, v_n\}$ such that $e_j=(v_{j-1},v_j)$ for all $j=1,\dots, n$.
If there is a path between two vertices $v$ and $w$ they are called
\emph{(path) connected}. This relation partitions the graph into \emph{(path
connected) components}. If a component contains a infinite number of
distinct vertices we call it an \emph{infinite component}.
The \emph{distance} between two vertices $v, w\in V$ is defined by
\begin{multline}
\dist_G (v,w):= \dist (v,w)
\\
:= \min \{\text{length of } p \mid  p \text{ is a path connecting $v$ and $w$} \}
\end{multline}
Note that the distance between $v$ and $w$ in a sub-graph of $G$ may be larger than their distance
in the original graph $G$.
The \emph{vertex degree}   $\deg(v)$ of a vertex $v\in V$ equals the number of
edges $e\in E$, such that $v$ is an endpoint of $e$.

Let $G$ and $G'$ be graphs.
A map $\phi\colon G\to G'$ is called a \emph{graph-map} or
\emph{graph-homomorphism}, if $\phi\colon V(G)\to V(G')$, $\phi\colon
E(G)\to E(G')$  and if for any $e=(v, w)\in E(G)$, the image $\phi(e)$ equals
$(\phi(v), \phi(w))$. A
graph-map $\phi\colon G\to G$ which has an inverse graph-map is called a
\emph{graph-automorphism} or simply \emph{automorphism} of $G$.
\medskip

Let $\Gamma$ be a group of graph-automorphism acting on a graph $X$.
It induces a projection map $\proj \colon X \to X / \Gamma$. We assume
that the quotient is a finite graph. This implies in particular that the
degree of the vertices in $V$ is uniformly bounded. We denote the smallest upper bound
by $\deg_+$.
Chose a vertex $[v]\in V(X/ \Gamma)$ and a representative $v\in [v]
\subset V(X)$. Starting form $v$, lift pathwise the vertices and edges of $X/ \Gamma$ to
obtain a connected set of vertices and edges $\tilde\cF\subset X$, such that $\proj|_{\tilde\cF}
\colon \tilde\cF \to X/ \Gamma$ is a bijective map. The set $\cF:= \tilde\cF \cup \{v \in V(X) \mid v \text{ is an endpoint of an edge in } \cF \}$ is a graph, which we call \emph{fundamental domain}. Note that $\proj|_\cF
\colon \cF \to X/ \Gamma$ is a graph-map, which is bijective on the set of edges, but not on the set of vertices.
\medskip

We construct a probability space $(\Omega, \cA, \PP)$ associated
to percolation on $X$. Let $\Omega= \times_{v\in V} \{0, \infty\}$
be equipped with the $\sigma$-algebra $\cA$ generated by finite dimensional cylinders sets.
Denote by $\PP$ a probability measure on $\Omega$ and assume that the measurable shift transformations
\[
\tau_\gamma\colon \Omega \to \Omega, \quad 
(\tau_\gamma \omega)_v = \omega_{\gamma^{-1}v}
\]
are measure preserving. Moreover, let the family $\tau_\gamma , \gamma \in \Gamma$
act ergodically on $\Omega$.
By the definition of $\tau_\gamma , \gamma \in \Gamma$
the stochastic field $q\colon \Omega\times V \to \{0,\infty\}$
given by $q(\omega,v)=\omega_v, v \in V$ is \emph{stationary} or \emph{equivariant},
i.e.{} $q(\tau_\gamma\omega,v)=q(\omega,\gamma^{-1}v) $.
An element $\omega$ of the probability space will be called
\emph{configuration}. The mathematical expectation associated to the probability $\PP$ will be denoted by $\EE$.

For a configuration $\omega$, a site $v$
with $q(\omega, v) =0$ will be called \emph{active} or \emph{undeleted}
and a site $v$ with $q(\omega, v) =\infty$ \emph{deleted}.
\smallskip

For each $\omega \in \Omega$ denote by
\[
V(\omega)= V(X(\omega))= \{v \in V \mid  q(\omega, v) = 0   \}
\]
the subset of active vertices, and denote by $X(\omega)$ the corresponding \emph{induced sub-graph} of $X$.
It is the sub-graph of $X$ whose vertex set is $V(\omega)$ and whose edge set is
\[
E(\omega)= E(X(\omega))= \{e \in E(X) \mid  \text{ both endpoints of $e$ are
in } V(\omega)   \}
\]

Let $\Lambda=(V(\Lambda), E(\Lambda))$ be an (deterministic) induced sub-graph of $X$.
It gives rise to a random family of induced sub-graphs $\Lambda(\omega):= X(\omega) \cap \Lambda$.
\smallskip

On any of the graphs introduced so far we will consider 
operators of finite hopping range. The easiest example to have in mind is the adjacency
operator  $A_\omega$ considered already in \eqref{e-QPM}.
More generally, a \emph{operator of finite hopping range} $H$ on a graph
$G$ is a linear map $H\colon l^2(V(G))\to l^2(V(G))$ such that there exists $C,
R\le\infty$ with
\begin{enumerate}[(i)]
\item $H(v,w) = H(w,v)$
\item $H(\gamma v,\gamma w) = H(v,w)$ for all  $\gamma \in \Gamma$
\item $|H(v,w)| \le C$ and
\item $H(v,w)=0$ if $\dist(v,w)\ge R$
\end{enumerate}
for all $v,w \in V(G)$. Here $H(v,w):= \la \delta_v, H \delta_w\ra$ and $\delta_v\in l^2(V(G))$
is the function  taking the value $1$ at $v$ and $0$ elsewhere.

For a sub-graph $G \subset X$ and a finite hopping range operator $H$ denote by
$H^G $ the compression of $H$ to $l^2(V(G))$, in other words
\[
H^G(v,w) =  H(v,w) \text{ if $ v,w \in G $ and  $H^G(v,w) = 0 $ otherwise}
\]
If $V=V(G)$ is finite, $H$ is a $(\nr{V}\times\nr{V})$-matrix, where
$\nr{\cdot}$ denotes the cardinality of a set. Thus the spectrum of $H^G$ is real and
consists entirely of eigenvalues $\lambda_i(H^G)$, which we enumerate in
increasing order counting multiplicity.
Let us define the normalized eigenvalue counting function of $H^G$ as
\[
N^G(H,\lambda)
:= \frac{\nr{\{i \in \NN\, \mid  \lambda_i(H^G) < \lambda \}}}{ \nr{V}}
\]
\smallskip

We assume that the discrete group $\Gamma$ is \emph{amenable}, i.e.~there
exists a {F\o lner} sequence $\{I_j\}_j$ of finite, non-empty subsets of
$\Gamma$. A sequence $\{I_j\}_j$ is called \emph{{F\o lner} sequence} if
for any finite $K\subset \Gamma$ and $\epsilon >0$
 \be
 \label{e-FS}
 \vert I_j \Delta K I_j \vert  \le \epsilon \,{\vert I_j\vert}
 \ee 
for all $ j$ large enough. Since the quotient $X / \Gamma $ is compact,
it follows that $K:= \{\gamma \in \Gamma \mid \gamma\cF \cap \cF \neq \emptyset \}$ is a finite generator set for $\Gamma$, cf.~\S 3 in \cite{AdachiS-1993} for a similar statement in the context of  manifolds. Now for finitely generated amenable groups there exists  a {F\o lner} sequence of subsets, 
which is increasing and exhausts $\Gamma$, cf.~Theorem 4 in \cite{Adachi-1993}. 
From \cite{Lindenstrauss-2001} we infer that each {F\o lner} sequence has an tempered
subsequence. A \emph{tempered} {F\o lner} sequence is a sequence
which satisfies in addition to \eqref{e-FS} the growth condition
\[
\text{ there exists $C < \infty$ such that for all } j \in \NN \ : \
\nr{I_{j} I_{j-1}^{-1}} \le C \nr{I_{j}}
\]
To each increasing, tempered {F\o lner} sequence associate an
\emph{admissible exhaustion} $\{\Lambda_j\}_j$ of $X$ given by
\[
\Lambda_j := \bigcup_{\gamma \in I_j^{-1}} \gamma \cF \subset X
\]
where $I_j^{-1}:= \{\gamma |\gamma^{-1}\in I_j \}$.
\smallskip

For a finite hopping range operator $H$, a  F\o lner sequence $\{I_j\}_j$, and a
random configuration $\omega\in \Omega$ introduce for brevity sake
the following notation:
$H_\omega:=H^{X(\omega)}$,
$H_\omega^j:=H^{\Lambda_j(\omega)}$,
and $N_\omega^j(\lambda):=N(H_\omega^j,\lambda)$.
Denote by $P_\omega (I):=\chi_I(H_\omega)$ the spectral projection of $H_\omega$ associated
to the energy interval $I$.

\begin{thm}
\label{t-exIDS}
There exist a distribution function $N$ called \emph{integrated density of
states} such that for almost all $\omega \in \Omega$ and any
admissible exhaustion $\Lambda_j , j \in \NN$ we have
\be
\label{t-d-IDS}
\lim_{j\to \infty} N_\omega^j (E) = N(E)
\ee
at all continuity points of $N$. The following \emph{trace formula} holds for the
IDS
\[
N(E) = \frac{1}{\nr{\cF}} \EE \left \{ \Tr ( \chi_{\cF} P_\omega
(]-\infty, E[))   \right \}
\]
\end{thm}

We say that the IDS $N$ is associated to the sequence of random operators
$\{H_\omega^j\}_{\omega \in \Omega}, j \in \NN$. Next we address the
question of boundary condition dependence. Denote $\Lambda^c= X\setminus
\Lambda$.

\begin{prp}
\label{p-bcIND}
Let $H$ be a finite hopping range  operator, $\Lambda_j,  j \in \NN$ an
admissible exhaustion and $\tR \in \NN, C < \infty$. Let $B^j\colon
l^2(\Lambda_j)\to l^2(\Lambda_j), j \in \NN$ be any sequence of symmetric operators
such that for all $v,w\in V$ we have  $|B^j(v,w)|\le C$ and $B^j(v,w)=0$
if $\dist(v, \Lambda_j^c)+ \dist(w, \Lambda_j^c) >\tR$. Then  the IDS'
associated to the sequences $\{H_\omega^j\}_{\omega\in \Omega}, j \in \NN$
and  $\{H_\omega^j + B^j \}_{\omega\in \Omega}, j \in \NN$ coincide.
\end{prp}

Next we establish the non-randomness of the spectrum of $H_\omega$ and its
components, its relation to the IDS and an understanding of the IDS as a
\emph{von Neumann trace}.
Denote by $\sigma_{disc}, \sigma_{ess}, \sigma_{ac}, \sigma_{sc}, \sigma_{pp}$
the discrete, essential, absolutely continuous, singular continuous, and pure point part of the spectrum.
Denote by $\sigma_{comp}$ the set of eigenvalues which posses an eigenfunction with compact, i.e.~finite, support.
In the following theorem $\Gamma$ need not be amenable, but $X$ must be countable.

\begin{thm}
\label{t-nrSpec}
There exists a $\Omega' \subset \Omega$ of full measure and subsets of the
real numbers
$\Sigma$ and $ \Sigma_\bullet$, where $\bullet \in\{disc, ess, ac, sc, pp, comp\}$,
such that  for all $\omega\in \Omega' $
 \[
  \sigma(H_\omega)=\Sigma \quad \text{ and } \quad \sigma_\bullet (H_\omega)= \Sigma_\bullet
 \]
for any $\bullet = disc, ess, ac, sc, pp, comp$. If $\Gamma$ is infinite,
$\Sigma_{disc}=\emptyset$. The almost-sure spectrum $\Sigma$ coincides with the set of points of
increase of the IDS
\[
\Sigma = \{ \lambda\in \RR \mid  N(\lambda+\epsilon) > N(\lambda
-\epsilon) \text{ for all $\epsilon >0$}\}
\]
Furthermore,  $N$ is the distribution function of the spectral measure of
the
direct integral operator
\[
\cH:= {\int_\Omega}^\oplus H_\omega \, d \PP(\omega)
\]
On the von Neumann algebra associated to $\cH$ there is a canonical trace
and $N(E)$ is the value of this trace on the spectral projection of $\cH$
associated to the interval $]-\infty, E[$.
\end{thm}

\section{Proofs of the theorems}
Let $H$ be a finite hopping range
 operator and assume without loss of generality $|H(v,w)| \le 1$
for all matrix elements. It follows that the $l^2$-norm of $H$ is bounded
by $K:=2 \deg_+^{R}$. Since $H$ is symmetric, it is a selfadjoint
operator. In particular the spectrum of $H_\omega$ is contained in $[-K,K]$ for all $\omega\in\Omega$.

Each $\gamma \in \Gamma$ induces an unitary operator $U_{\omega,\gamma} \colon l^2(V(\tau_\gamma^{-1}\omega)) \to l^2(V(\omega))$, $(U_{\omega,\gamma} f)(v):= f(\gamma^{-1}v)$. Note that $V(\tau_\gamma\omega)=\gamma V(\omega)$. By the definition  of $\tau_\gamma$ the
action of $\Gamma$ on $\Omega$ and on $X$ is compatible:
\be
\label{e-equivar}
U_{\omega,\gamma} H_\omega U_{\omega,\gamma}^* = H_{\tau_\gamma \omega}
\ee

The equivariance formula \eqref{e-equivar} implies
\be
\label{e-equivarf}
U_{\omega,\gamma} f(H_\omega) U_{\omega,\gamma}^* = f(H_{\tau_\gamma \omega})
\ee
for any polynomial $f$. For continuous functions $ f, g$ we have 
$\| f(H_\omega)- g(H_\omega)\| \le \|f-g \|_\infty$.
Thus $f_n\to f$ in $C([-K,K],\|\cdot\|_\infty)$ implies $f_n(H_\omega)\to f(H_\omega)$ in operator norm, and \eqref{e-equivarf} extends by Weierstra\ss' approximation theorem
to all $f \in C([-K,K])$. By taking scalar products we obtain the corresponding equivariance relation for the matrix elements:
\[
f(H_\omega) (\gamma^{-1}v,\gamma^{-1}w) = f(H_{\tau_\gamma \omega}) (v,w)
\]
\smallskip

For the proof of the main Theorem \ref{t-exIDS} we need two key
ingredients: an estimate of boundary effects on traces and a sufficiently
general ergodic theorem, which will be applied to trace functionals of the type
\[
F(\omega)
:= \nr{\cF}^{-1} \sum_{v \in \cF} f(H_{\omega}) (v,v) 
=  \nr{\cF}^{-1}\Tr \left(f(H_{\omega}) \chi_{\cF}   \right)
\]

Let us first estimate the boundary effects.

\begin{prp}
\label{p-traceapp}
Let $f(x) = x^m$ for $m \in  \NN$. Then
\[
\sup_{\omega \in \Omega}\frac{1}{\nr{\Lambda_j}} \left |
\Tr(f(H_\omega^j))- \Tr(\chi_{\Lambda_j}f(H_\omega)) \right|
\to 0
\]
as $j \to \infty$.
\end{prp}
\begin{proof}
We introduce the notion of a thickened boundary on a graph. For a sub-graph $\Lambda$ and $h \in \NN$
set $\partial_h \Lambda:= \{v \in \Lambda \mid \dist(v, \Lambda^c) \le h \}$.
We expand the trace of powers of $H_\omega^j$:
\[
\Tr (H_\omega^j)^m
= \sum_{v\in \Lambda_j} (H_\omega^j)^m (v,v)
= \sum_{v\in \Lambda_j} \sum_{v_1,\dots, v_{m-1}\in \Lambda_j}
H_\omega(v,v_1) \dots H_\omega(v_{m-1},v)
\]
By an analogous formula for $\Tr(\chi_{\Lambda_j} H_\omega^m)$ we obtain
\bea
\Tr [\chi_{\Lambda_j} H_\omega^m-  (H_\omega^j)^m]
= \sum_{v\in \Lambda_j} \sum_{\bullet}  H_\omega(v,v_1) \dots
H_\omega(v_{m-1},v)
\\
\le
\nr{\partial_{ R m}\Lambda_j} \deg_+^{m^2 R}
\eea
where the bullet denotes summation over $m-1$-tuples (paths) in $V(X)$
with at least one vertex outside $\Lambda_j$. By the F\o lner property of
the sequence $I_j, j \in \NN$ we have:
\[
\lim_{j \to \infty}\frac{\nr{\partial_{ h}\Lambda_j} }{\nr{\Lambda_j}}=0
\quad \text{  for any } h \ge 0
\]
This is the content of Lemma 2.4 in \cite{PeyerimhoffV-2002}. In fact,
there manifolds are considered, but the proof applies literally to the
case of graphs.
\end{proof}

Lindenstrauss proved in \cite{Lindenstrauss-2001} a remarkable ergodic
theorem which
applies to locally compact, second countable, amenable groups. It includes
the following statement for discrete groups.

\begin{thm}
\label{ergthm}
Let $\Gamma$ be an amenable discrete group and $(\Omega,\cB_\Omega,\PP)$
be a probability space.
Assume that $\Gamma$ acts ergodically on $\Omega$ by measure preserving
transformations $\tau_\gamma$.
Let $\{ I_j \}_j $ be a tempered {F\o lner} sequence in $\Gamma$.
Then for every $F \in L^1(\Omega)$
\begin {equation}
\label{aver}
\lim_{j \to \infty} \frac{1}{\vert I_j \vert} \sum_{\gamma \in I_j}
F(\tau_\gamma \omega)
= \EE\{F\}
\end{equation}
for almost all $\omega \in \Omega$.
\end{thm}
In the application we have in mind $F\in L^{\infty}$, so the convergence
holds in the $L^1$-topology, too. 
\smallskip

\begin{proof}[Proof of Theorem \ref{t-exIDS}]
To prove the claimed convergence of distribution functions it is sufficient to establish $ \int f dN_\omega^j \to \int f dN$ for all $f \in C([-K,K])$. This can in turn be reduced to proving the convergence in the case where $f$ is a polynomial: By Weierstra\ss' approximation theorem the polynomials are dense in $C([-K,K])$. Let $f_k$ be a sequence of polynomials such that $\|f_k -f\|_\infty \to 0 $. For any $\epsilon >0$ one can choose $k$ large enough such that  $\|f_k -f\|_\infty < \epsilon/3$ and subsequently $j$ large enough (depending on $\epsilon$ and $f_k$) such that   $|\int f_k dN_\omega^j - \int f_k dN|< \epsilon/3$. 
Since the total measure of $N$ and any $N_\omega^j$ is bounded by one,
\begin{multline*}
\l|\int f dN_\omega^j - \int f dN \r|  
\\
\le 
\l|\int (f  -  f_k) dN_\omega^j\r| 
+ \l|\int f_k dN_\omega^j - \int f_k dN \r| 
+ \l|\int (f_k  -f) dN \r|
< \epsilon
\end{multline*}
 Thus it is sufficient to prove the convergence of moments
\be
\label{e-mc}
\lim_{j \to \infty} \int_\RR \lambda^m N_\omega^j(d\lambda) = \int_\RR
\lambda^m N(d\lambda)
\quad \text{ for all $m \in \NN$}
\ee
\\
Next we show that the limit on the left hand side equals
$\frac{1}{\nr{\cF}}\EE \left \{ \Tr (\chi_\cF H_\omega^m)  \right \} $
for almost all $\omega$. For this aim we write the
moment of the IDS as a trace
\bea
\int_\RR \lambda^m N_\omega^j(d\lambda) =\nr{\Lambda_j}^{-1} \Tr(f(H_\omega^j))
\eea
which by Proposition \ref{p-traceapp} converges  for $j \to \infty$ to the same limit as
$ \nr{\Lambda_j}^{-1} \Tr(\chi_{\Lambda_j}f(H_\omega))$.
Now we decompose the trace according to local contributions and apply Lindenstrauss' theorem
\begin{multline*}
\nr{\Lambda_j}^{-1} \Tr(\chi_{\Lambda_j}f(H_\omega))
=
\nr{ \Lambda_j}^{-1} \sum_{v \in \Lambda_j} f(H_\omega^j) (v,v)=
\\
\nr{ \Lambda_j}^{-1} \sum_{\gamma \in I_j} \sum_{v \in \cF} f(H_\omega)
(\gamma^{-1}v,\gamma^{-1}v)
=
\nr{ I_j}^{-1} \sum_{\gamma \in I_j} \nr{\cF}^{-1}\sum_{v \in \cF}
f(H_{\tau_\gamma\omega}) (v,v)
\\
\to \EE\{F\} \quad \text{ as $j\to \infty$ for almost all $\omega\in
\Omega$}
\end{multline*}
where $F(\omega)=\nr{\cF}^{-1}\sum_{v \in \cF} f(H_\omega) (v,v) =
\nr{\cF}^{-1}\Tr(\chi_{\cF}f(H_\omega)) $.
Set \\ $E_\omega(\lambda) := P_\omega(]-\infty, \lambda[)$.
The expectation of $F$  equals
\begin{multline*}
\frac{1}{\nr{\cF}}\EE  \left\{\sum_{v \in \cF} \int \lambda^m E_\omega(d\lambda)(v,v) \right\}
\\
=\frac{1}{\nr{\cF}}\int \lambda^m  \EE  \left\{\Tr (\chi_{\cF} E_\omega(d\lambda) ) \right\}
= \int \lambda^m N(d\lambda)
\end{multline*}
\end{proof}

\begin{proof}[Proof of Proposition \ref{p-bcIND}]
Denote $\tilde H_\omega^j := H_\omega^j  + B^j$. Similarly as in the proof of
Proposition \ref{p-traceapp} we have
\begin{multline*}
\Tr [(\tilde H_\omega^j)^m -  (H_\omega^j)^m]
\\
= \sum_{v\in \Lambda_j} \sum_{\bullet}
H_\omega(v,v_1) \dots H_\omega(v_{m-1},v) -\tilde H_\omega(v,v_1) \dots \tilde H_\omega(v_{m-1},v)
\\
\le
[1+(1+ C)^m]\, \nr{\partial_{ R m+ \tR}\Lambda_j} \, \deg_+^{m^2 r}
\end{multline*}
Here the bullet denotes the summation over all paths
$v_1,\dots, v_{m-1}\in \Lambda_j$ with at least one vertex in
$\partial_{\tR}\Lambda_j$ and $r := \max(R, \tR)$. By the F\o lner
property
$\nr{\Lambda_j}^{-1} \Tr [(\tilde H_\omega^j)^m -  (H_\omega^j)^m]$ converges
to zero as $j \to \infty$.
\end{proof}
\medskip

\begin{proof}[Proof of Theorem \ref{t-nrSpec}]
First we prove the non-randomness of $\sigma_{comp}$. Set
\begin{align}
\label{e-tildeSigma}
\tilde\Sigma := & \{ E \in \RR \mid \exists \text{ finite induced sub-graph } G\subset X \text{ and } f \in l^2(G)
\\ \nonumber
&\text{ such that } H^G f=Ef\}
\end{align}
Then $\sigma_{comp}(H_\omega)\subset \tilde\Sigma$ for all $\omega \in \Omega$. Since $X$ is countable,
there exists a countable exhaustion  of $X$ by finite sets $D_j, j \in \NN$. If we set
\[
\tilde\Sigma_j := \{ E \in \RR \mid \exists \omega \in \Omega \text{ and } f \in l^2(D_j(\omega)) \text{ s.t.{} } H^{D_j(\omega)} f=Ef\} ,
\] 
then $\tilde\Sigma= \cup_j \tilde\Sigma_j$ and thus $\tilde\Sigma$ is countable.

For any $E \in \tilde\Sigma $ set $\Omega_E:= \{\omega \mid \exists f \text{ with finite support and } H_\omega f=Ef \}$. This set is invariant under the ergodic action of $\Gamma$ by the transformations $\tau_\gamma$. Therefore,
either $\PP(\Omega_E)=1$ or $\PP(\Omega_E)=0$. In the first case set $\tilde\Omega_E=\Omega_E$ and $E \in \Sigma_{comp}$, in the second set $\tilde\Omega_E=\Omega_E^c$ and $E \not\in \Sigma_{comp}$. Here the superscript $^c$ denotes the complement of a set. The set $\tilde\Omega :=\cap_{E \in \tilde\Sigma} \, \tilde\Omega_E$ has full measure
and each $\omega \in \tilde \Omega$ satisfies $\sigma_{comp}(H_\omega)=\Sigma_{comp}$.

The remaining statements of the theorem
follow from the results of \cite{LenzPV-2002?}.
One just has to check that the required assumptions are satisfied. This is
not hard, but it would require to introduce the notion of grupoids and
basic features of Connes' non-commutative integration theory
\cite{Connes-1979}. Therefore we leave the details of the proof of Theorem
\ref{t-nrSpec} to another occasion.
\end{proof}

\section{Outlook: finitely supported and exponentially decaying states}
\label{s-disc}
Once we have a rigorous definition of the integrated density of states for
the QPM, we can study finer spectral properties.
One of the main interest in the physics literature are the properties of
bound states, and their contribution to numerically
observed "peaks" of the \emph{density of states}. This quantity is the distributional derivative of the IDS.
In the following we restrict
our discussion to the QPM corresponding to the adjacency operator $A$ on
the lattice $\ZZ^d$, and to the sequence $q(\cdot, v), v \in \ZZ^d$, consisting of independent, identically distributed random variables.
\medskip

There seem to be three different types of bound states of the QPM: finite
cluster states, molecular states and exponentially decaying states,
cf.~\cite{deGennesLM-59a,deGennesLM-59b,KirkpatrickE-72,ShapirAH-82,ChayesCFJS-86}.
Finite cluster states occur since almost surely there are active clusters
of finite size, which, consequently, can support only bound states. These
states are mathematically not challenging.
\\

However, as pointed out earlier, there exist molecular states. They are
eigenvectors of the adjacency operator restricted to the active sites with
support on a finite region of  the infinite cluster. This means that due
to the deletion of sites  the unique continuation property of eigenfunctions of the
adjacency operator on the lattice breaks down.

For the analysis of molecular states it is convenient to introduce the
restriction $A_\omega^\infty$ of $A$ to the infinite active cluster $X^\infty(\omega)$,
and to define the corresponding IDS by
\[
N^\infty(\lambda)
= \nr{\cF}^{-1} \EE \{ \Tr [ \chi_{\cF^\infty(\omega)} E_\omega^\infty (\lambda)]\}
\]
where $\cF^\infty(\omega) = \cF\cap X^\infty(\omega)$ and $\cF = \{0\}$. $N^\infty$ is
self-averaging, i.e.~can be defined by an exhaustion procedure, similarly as $N$ in Theorem \ref{t-exIDS}.
\smallskip

Here is a result on molecular states, whose proof is given in \cite{Veselic-QP2}.

\begin{thm}
The set of points of discontinuity of the IDS $N$ of $\{A_\omega\}_\omega$
coincides with the set of points of discontinuity of the IDS $N^\infty$ of
$\{A_\omega^\infty\}_\omega$ and equals $\tilde\Sigma $ as defined in \eqref{e-tildeSigma}.
\end{thm}

Finally, one can hope that a multi-scale argument will yield a proof of the existence of exponentially decaying
eigenstates, as in the case of the Anderson model \cite{Anderson-58,FroehlichS-83,DreifusK-89,Stollmann-01}. For this aim, one has to develop new tools to deal with the
singular randomness present in the QPM. On the other hand, once this is
done, one might use the new ideas to approach the exponential localization problem for
the Anderson model with Bernoulli disorder of the coupling constants. In
the multi-dimensional case this
is a problem which is open since decades.

\def\cprime{$'$}

\end{document}